\documentstyle[twocolumn,aps,floats,psfig]{revtex}
\begin{document}
\author{Mauricio Cataldo$^a$
{\thanks{E-mail address: mcataldo@alihuen.ciencias.ubiobio.cl}},
Norman Cruz$^b$ {\thanks{E-mail address: ncruz@lauca.usach.cl}},
Sergio del Campo$^c$ {\thanks{E-mail address: sdelcamp@ucv.cl}}
and Alberto Garc\'\i a $^d$ \thanks{E-mail address:
alberto.garcia@fis.cinvestav.mx} }
\address{
$^a$ Departamento de F\'\i sica, Facultad de Ciencias, Universidad
del B\'\i o-B\'\i o, Avda. Collao 1202, Casilla 5-C, Concepc\'\i
on, Chile \\ $^b$ Departamento de F\'\i sica, Facultad de Ciencia,
Universidad de Santiago de Chile, Avda. Ecuador 3493, Casilla 307,
Santiago, Chile. \\ $^c$ Instituto de F\'\i sica, Facultad de
Ciencias B\'asicas y Matem\'aticas, Universidad Cat\'olica de
Valpara\'\i so, Avda. Brasil 2950, Valpara\'\i so, Chile \\ $^d$
Departamento de F\'\i sica, CINVESTAV-IPN, Apartado Postal 14-740,
C.P. 07000, M\'exico, D.F. M\'exico
\\ \smallskip\ }
\title{(2+1)-Dimensional Black Hole with Coulomb-like Field.}
\maketitle
\begin{abstract}
{\bf Abstract:}{A (2+1)-static black hole solution with a
nonlinear electric field is derived.  The source to the Einstein
equations is a nonlinear electrodynamics, satisfying the weak
energy conditions, and it is such that the energy momentum tensor
is traceless. The obtained solution is singular at the origin of
coordinates. The derived electric field $E(r)$ is given by
$E(r)=q/r^2$, thus it has the Coulomb form of a point charge in
the Minkowski spacetime. This solution describes charged
(anti)--de Sitter spaces. An interesting asymptotically flat
solution arises for $\Lambda=0$. \\} {Keywords: 2+1 dimensions,
Non-Linear black hole }\\ PACS numbers: 04.20.Jb
\end{abstract}
\smallskip\

Most of the papers on 2+1-solutions coupled to electromagnetic
fields are done for Maxwell
electrodynamics~\cite{GoSiAl,Cataldo1}, i.e. the electromagnetic
tensor is derived, in 2+1 theory, from a Lagrangian which is
proportional to the single invariant, namely $L \propto F$, with
$F=\frac{1}{4} F_{\mu \nu } F^{\mu \nu }$. On the other hand, the
introduction of electrodynamics of nonlinear
type~\cite{AyGa,SaGaPl,GiRa} , where the dependence on the
invariant is enlarged, has
proved to be fruitful~\cite{Cataldo2}.

In 3+1 electromagnetism, the Maxwell energy momentum tensor is
given by  $T_{\mu \nu }=  \left (F_{\mu \alpha}
F^{\alpha}_{\nu}-g_{\mu \nu} F \right)/ (4 \pi)$ and consequently
is trace free. If one constructs a nonlinear electrodynamics based
on the invariant $F$, i.e. $L=L(F)$, the resulting energy momentum
tensor is given as $T_{\mu \nu}=g_ {\mu \nu} L_{F}-F_{\mu
\alpha}F^{\alpha}_{\nu} L$ and its trace becomes $T=4 L_{F}-4 F
L$. Thus, if one demands that the trace to be vanished, i.e.
$T=0$, we find that $L_{F}-F L=0$, whose solution is $L \propto
F$. In Minkowski spacetime the Maxwell theory is singled out among
all nonlinear theories by the vanishing of the trace. Recall that
in 3+1 there is a second invariant -- a pseudoscalar
$\breve{G}=\frac{1}{4} \overstar{F}_{\mu \nu} F^{\mu \nu}$, where
$\star$ stands for the duality operation. We may include into the
Lagrangian for building up a wider nonlinear theories in which the
Born-Infeld electrodynamics\cite{BoIn}is an example.

In 2+1 spacetime, the Maxwell energy momentum tensor is of the
same form as in 3+1 dimensions, but the trace contrary of the 3+1
case occurs to be non-vanishing, i.e. $T=F/4 \pi \neq 0$. Hence,
this 2+1 Maxwell theory has always trace. The electric field for a
static circularly symmetric  metric coupled to a Maxwell field
occurs to be proportional to the inverse of $r$, i.e. $E(r)
\propto 1/r$, hence the potential $A_0$ is  logarithmic, i.e. $A_0
\propto \ln r$, and consequently blows up at $r=0$ and $r$ going
to infinity.

In this paper we are interested in electromagnetic theories in
which the energy momentum tensor is traceless. This condition
restricts the class of nonlinear electrodynamics to be studied.
Incidentally the traceless non-linear electrodynamics, in 2+1
dimensions, occurs to be unique with Lagrangian proportional to
$F^{3/4}$. Restricting our study to a static circularly symmetric
metric, the resulting electromagnetic field for this theory
surprisingly is proportional to the inverse of $r^2$, i.e. a
Coulomb law for a point charge in 3+1 Minkowski space. Moreover,
the energy momentum tensor fulfills the weak energy conditions.
The resulting metric depends on three parameters:the mass, $M$,
the cosmological constant, $\Lambda$, and the charge, $q$.  When
$\Lambda < 0$, for certain range of values of these constants, it
describes black holes , i.e. charged anti-de Sitter spacetimes
with inner and outer horizons. These horizons are roots of an
algebraic cubic equation. In this family, when the horizons shrink
to a single one, $r_{extr}$, one obtains an extreme black hole. If
these roots are complex we obtain a naked singular solution. For
$\Lambda >0$ there is only one positive root and the corresponding
gravitational field describes a solution with a cosmological
horizon. A detailed analysis depending on the values of the
constants is given. The behavior of the $g^{r r}$ is plotted for
different branches. It is worthwhile to point out that this metric
allows for asymptotically flat solution for vanishing $\Lambda$.
Finally, thermodynamics aspects for the studied metric are
explicitly worked out.

We are using electromagnetic Lagrangian depending upon a single
invariant $L(F)$, where $F= 1/4 F^{a b} F_{ab}$. To deal with
physically reasonable theories, the fulfillment of the weak energy
conditions is imposed on the corresponding energy momentum tensor:
for any timelike vector $u^a$, $u^a u_a= -1$ (we are using
signature -- + +) one requires $T_{a b} u^a u^b \geq 0$ and $q_a
q^a \leq 0$, where $q^a= T^a _b u^b$. The action of the
(2+1)-Einstein theory coupled with nonlinear electrodynamics is
given by
\begin{eqnarray}
\label{action} S=\int \sqrt{-g} \left(\frac 1{16\pi} (R-2\Lambda)
+ L(F) \right) \,d^3x,
\end{eqnarray}
with arbitrary, at this stage, the electromagnetic Lagrangian
$L(F)$.   We are using units in which $c=G=1$. Since there is an T
ambiguity in the definition of the gravitational constant (there
is not Newtonian gravitational limit in (2+1)-dimensions) one can
maintain the factor $1/16 \pi$ in the action to keep the
parallelism with (3+1)-gravity. The variation with respect to the
metric gives the Einstein equations
\begin{eqnarray}
G_{ab} + \Lambda g_{a b}= 8 \pi T_{ab},
\end{eqnarray}
\begin{eqnarray}
\label{tensor electromagnetico}
T_{ab}=  g_{ab} L(F)- F_{ac} F_{b}^{\,\,c} L_{_{,F}} ,
\end{eqnarray}
while the variation with respect to the electromagnetic potential
$A_{a}$ entering in $F_{ab}= A_{b,a} - A_{a,b}$, yields the
electromagnetic field equations
\begin{eqnarray}
\label{poisson}
\nabla_{a} \left( F^{ab} L_{_{,F}}  \right)=0,
\end{eqnarray}
where $L_{_{,F}}$ denotes the derivative of $L(F)$ with respect to
$F$. In what follows we shall restrict ourselves to the study of
the nonlinear field such that the energy momentum
tensor~(\ref{tensor electromagnetico}) has vanishing trace. The
trace of this tensor(\ref{tensor electromagnetico}) occurs to be
\begin{eqnarray} T = T_{a b} g^{a b}= 3 L(F) - 4 F L_{_{,F}},
\end{eqnarray}
thus, if one requires $T$ to vanish, one obtains
\begin{eqnarray}
\label{Lagrangiano}
L = C \, |F|^{3/4},
\end{eqnarray}
where $C$ is a constant of integration, and bars denote absolute
value. One can rewrite this Lagrangian as
\begin{eqnarray}
\label{culo} L = C \, \left |\frac{1}{2} \left (B^2 - E^2  \right
) \right|^{3/4},
\end{eqnarray}
when referred to orthonormal local lorentzian basis.

In what follows, we shall look for black hole solutions in the
static case. The metric we are dealing with is given by
\begin{eqnarray}
\label{metrica}
ds^{2}= - f(r) dt^{2} + \frac{dr^{2}}{f(r)} + r^{2} d \Omega^{2},
\end{eqnarray}
where $f(r)$ is an unknown function of the variable r. Next, we
assume that the electromagnetic field is given by: the electric
field $E_{r}= F_{tr}$ and the magnetic scalar field $B:=F_{\phi
r}$. The corresponding Maxwell equations are
\begin{eqnarray}
\label{rit710}
{\frac{d}{dr}}[r E L_{,F}]=0,\,\,\,{\frac{d}{dr}} \left [\frac{f}{r}B
L_{,F} \right ]=0.
\end{eqnarray}
By virtue of the Einstein equations the $B$ field has to vanish.
In fact,  the Ricci tensor components, evaluated for the metric
(\ref{metrica}), yield the following relation
\begin{eqnarray}
\label{rit11}  A := R_{tt} +f^2 R_{rr}=0,
\end{eqnarray}
while the evaluation of the same relation using the
electromagnetic energy-momentum gives
\begin{eqnarray}
A=-8 \pi L_{,F}(\frac{f}{r}B)^2.
\end{eqnarray}
Therefore, the scalar magnetic field $B$ has to be equated to
zero, $B=0$. Hence, for the metric (\ref{metrica}), the only case
allowed is just the one with the electric field $E$. Consequently,
the electromagnetic field tensor can be given as
\begin{eqnarray}
\label{tensorr}
F_{a b} = E(r) \left ( \delta^{t }_{a }  \delta^{r}_{ b} - \delta^{r }_{a}
\delta^{t}_{b} \right ).
\end{eqnarray}
The invariant $F$ occurs to be
\begin{eqnarray}
\label{invariante}
2 F = -E^{2}(r).
\end{eqnarray}
Integrating the equation (\ref{rit710}) for $E$, one obtains
\begin{eqnarray}
\label{Elec}
 E(r) L_{,F}= - \frac{q}{4 \pi r}.
\end{eqnarray}
where $q$ is an integration constant.
Using now~(\ref{invariante}) we express the derivative $L_{,F}$ as function of
$r$, as follows
\begin{eqnarray}
\label{Elec2}
 L_{,r}= \frac{q}{4 \pi r} E_{,r}.
\end{eqnarray}
From~(\ref{culo}), setting $B=0$, we arrive at $L = C \, E^{3/2}$.
Entering this $L$ into~(\ref{Elec2}) one gets  $E =(q^2/6 \pi C)^2
1/r^2$. Choosing now $C= \sqrt{|q|}/6 \pi$, the electric field
becomes
\begin{eqnarray}
E(r)= \frac{q}{r^2},
\end{eqnarray}
which coincides with the standard Coulomb field for a point charge
of the Maxwell theory in the four dimensional Minkowski space. The
Lagrangian in this case is given by
\begin{eqnarray}
L = \frac{q^2}{6 \pi r^3}= \frac{\sqrt{|q|}}{6 \pi} E^{3/2}.
\end{eqnarray}
It is easy to check up that the energy momentum tensor for this
Lagrangian  satisfies the weak energy conditions:  $q_a q^a \leq
0$, where $q^a= T^a _b u^b$ for any timelike vector $u^a$, which
in terms of the related quantities is equivalent to the inequality
\begin{eqnarray}
- (L + E^2 L_{,F})= \frac{q^2}{12 \pi r^3} \geq 0.
\end{eqnarray}
Having established the electric field, we are now ready to search
for solutions of the Einstein equations,  which  equivalently are
\begin{eqnarray}
R_{a b} = 8 \pi T_{a b} + 2 \Lambda g_{a b},
\end{eqnarray}
where it has been considered that the trace of~(\ref{tensor
electromagnetico}) is equal to zero, i.e, $T=0$. These equations,
for $R_{tt} ( =- f^2 R_{_{r r}}) $ and $R_{_{\Omega \Omega}}$
components, yield respectively the equations:
\begin{eqnarray}
\label{tt}
f_{,rr}+ \frac{f_{,r}}{r}= -2 \Lambda + \frac{2 q^2}{3 r^3},
\end{eqnarray}
\begin{eqnarray}
\label{omegaomega}
f_{,r}= -2 \Lambda r - \frac{4 q^2}{3 r^2}.
\end{eqnarray}
It is easy to show that equation~(\ref{tt}), by virtue of the
equation~(\ref{omegaomega}) is just an identity. Therefore, the
only Einstein equation to be integrated is~(\ref{omegaomega}),
which gives
\begin{eqnarray}
\label{solucion penultimaaa} f(r)= D - \Lambda r^{2} + \frac{4
q^2}{3 r},
\end{eqnarray}
where $D$ is a constant of integration. We will see now that the
constant $D$ can be expressed in terms of the mass at infinity. To
demonstrate this, we shall use the quasilocal formalism developed
by Brown et al.~\cite{Brown1,Brown2} to evaluate the quasilocal
energy and mass of a stationary and axisymmetric asymptotically
non-flat spacetime. For the circularly symmetric
metric~(\ref{metrica}) the quasilocal energy ${\cal{E}}(r)$  and
the quasilocal mass $M(r)$ at a radial boundary $r$ can be shown
to be respectively
\begin{eqnarray}
{\cal{E}}= 2 (\sqrt{f_{0}(r)}- \sqrt{f(r)}),
\end{eqnarray}
\begin{eqnarray}
M(r)= {\cal{E}}(r) \sqrt{f(r)},
\end{eqnarray}
where $f_{0}(r) = g_{0}^{rr}(r)$ is a background metric function
which determines the zero of the energy. The function $f_{0}(r)$
can be obtained simply by assigning some special values to the
constants of integration, determining this way the reference
space-time. We set in~(\ref{solucion penultimaaa}) $q=D=0$,
arriving at the background space, which corresponds to an
asymptotic anti-de Sitter spacetime. The same background function
was used in~\cite{Brown2,Chan} for analogous calculations. For
$\Lambda=-1/l^2 < 0$, we get $\sqrt{f_{0}(r)}= r/l$ and then the
quasilocal energy and the quasilocal mass are given respectively
by
\begin{eqnarray}
\label{energia} {\cal{E}}(r) = \frac{r}{l} - \sqrt{D
+\frac{r^2}{l^2} + \frac{4}{3} \frac{q^2}{r}},
\end{eqnarray}
\begin{eqnarray}
\label{masa}
M(r) = 2 \frac{r}{l} \, \sqrt{D +\frac{r^2}{l^2} + \frac{4}{3} \frac{q^2}{r}}
  \nonumber \\
- 2 \left (D +\frac{r^2}{l^2} + \frac{4}{3} \frac{q^2}{r}  \right).
\end{eqnarray}
As $r \longrightarrow \infty$, the analogous ADM mass is defined
to be $M =: M(\infty)$. In our case we see from~(\ref{energia})
and~(\ref{masa}) that ${\cal{E}}(\infty)$ vanishes and
$M(\infty)=: M = -D$ correspondingly. The constant $D$ can be
thought of as the asymptotic observable mass, $M>0$, and we can
write
\begin{eqnarray}
\label{solucion penultima} f(r)= -M +\frac{r^{2}}{l^2} + \frac{4
q^2}{3 r}.
\end{eqnarray}

The case $\Lambda>0$ corresponds to an asymptotically de-Sitter
spacetime.

For vanishing cosmological constant, i.e. $\Lambda=0$, one has an
asymptotically flat solution coupled with a Coulomb-like field.

To establish the existence of  horizons, one has to require the
vanishing of the $g^{rr}$ component, i.e., $f(r)=-M -\Lambda r^{2}
+ \frac{4 q^2}{3 r} = 0$, for handling all possible cases. The
roots of this equation are
\begin{eqnarray}
r_{1} = \frac{h}{3 \Lambda} - \frac{M}{h},
\end{eqnarray}
\begin{eqnarray}
r_{2} = - \frac{h}{6 \Lambda} + \frac{M}{2 h} + \frac{i \sqrt{3}}{2} \,
\left ( \frac{h}{3 \Lambda} + \frac{M}{h} \right),
\end{eqnarray}
\begin{eqnarray}
r_{3} = - \frac{h}{6 \Lambda} + \frac{M}{ 2 h} - \frac{i \sqrt{3}}{2} \,
\left( \frac{h}{3 \Lambda} + \frac{M}{h} \right),
\end{eqnarray}
where
\begin{eqnarray}
h = \left[ \left (18 q^2+ 3 \sqrt{3 \left ( \frac{M^3 + 12 q^4
\Lambda}{\Lambda} \right) } \right) \Lambda^2 \right ]^{1/3}.
\end{eqnarray}
These equations give the location of the horizons (if there are
any). Since the coordinate $r$ range  from $0$ to infinity, we
exclude the negative roots. As it is well known from the
properties of cubic equations, the complex or real character of
the roots depends crucially on the sign of the radical $\alpha :=
M^3 / \Lambda + 12 q^4$. Due to $M>0$, the complex or real
character of these roots depends on values of $\Lambda$. For
$\alpha>0$, or $\Lambda
>-M^3/ 12  q^4 $, we get one real and two complex roots.
A more detailed study shows that the real root is $r_{1}$ and
positive. The remaining roots, $r_{2}$ and $r_{3}$, are complex.

\begin{figure}[ht]
\centerline{ \psfig{file=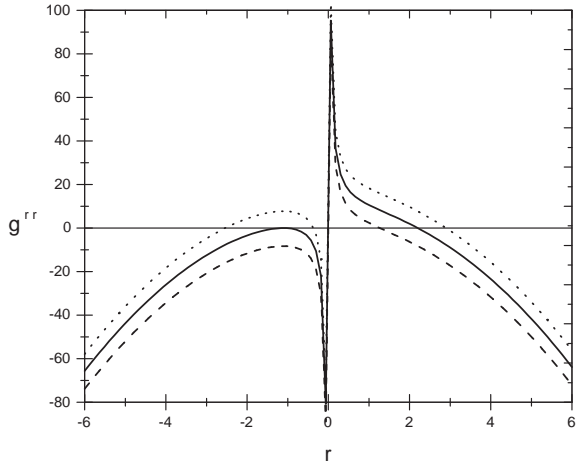,width=12cm,angle=0} }
\caption{Behavior of $g^{rr}$ for different values of $M$ and
$\Lambda >0$. The continuous line shows the behavior of the
extreme case, $M_{extr}=-7.268$, $\Lambda=2$ and $q=2$.  For the
same values of $\Lambda$ and $q$, the dashed and the dotted lines
show the behavior for $M=-15$ and $M=1$, respectively.}
\label{fig:flat01}
\end{figure}

Notice that $f(r)$ decreases for $r>0$, as we can see  from
FIG.~(\ref{fig:flat01}). In this plot we have taken the value
$M=1$ (we have also included negative values of $M$ for showing
the different behavior of the roots). Notice also that we have a
cosmological horizon.

\begin{figure}[ht]
\centerline{\psfig{file=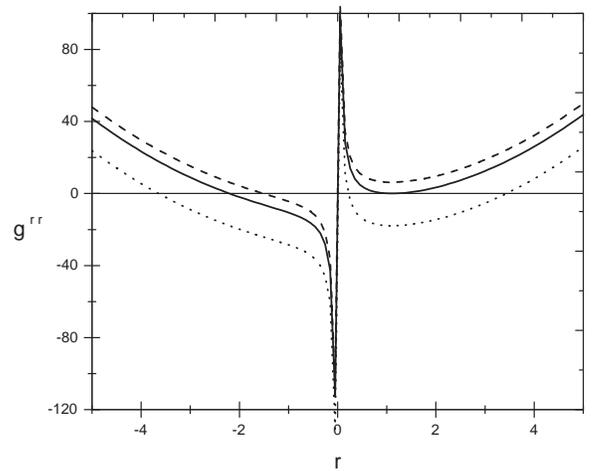,width=12cm,angle=0} }
\caption{Behavior of $g^{rr}$ for different values of $M$ and for
$\Lambda < 0$. The continuous line shows the behavior of the
extreme case, $M_{extr}=7.268$, $\Lambda=-2$ and $q=2$.  For the
same values of $\Lambda$ and $q$, the dashed and the dotted lines
show the behavior for $M=1$ and $M=25.15$, respectively.}
\label{fig:flat02}
\end{figure}

For $\alpha< 0 $ (or $\Lambda < -M^3/ 12  q^4$) there are three
real roots.  On the other hand, if $\Lambda < 0$ the corresponding
roots $r_{2}$ and $r_{3}$ can be complex or real. More exactly,
for $-M^{3}/12 q^{4}< \Lambda < 0$ we see that the roots $r_{2}$
and $r_{3}$ are complex. For $ \Lambda < - M^3/12 q^4$ we have
real roots only. In this case we may write $r_{2}$ and $r_{3}$ as
\begin{eqnarray}
\label{reales} r_{\pm} = -2 \sqrt{\frac{-M}{3 \Lambda}} \, \cos \,
\left ( \frac{1}{3} \, \arccos \left [ \frac{2 q^2}{\sqrt{-
\frac{M^3}{3 \Lambda}}} \right ]  \pm \frac{2 \pi}{3}\right ).
\end{eqnarray}
These roots represent the horizons of a black hole. The outer
horizon (the event horizon) $r_{+}$ and the inner horizon $r_{-}$.
If $\alpha=0$, or $\Lambda= -M^3/ 12  q^4$, then  it is obtained
an extreme black hole. In such case the mass becomes $M =
M_{extr}=  -(12 q^4 \Lambda )^{1/3}$ and the roots  are
\begin{eqnarray}
\label{rextr}
r_{1} = 2 \left( \frac{2 q^2}{3 \Lambda} \right )^{1/3}, \,\,\,
r_{2} = r_{3} = r_{extr} = - \left( \frac{2 q^2}{3 \Lambda} \right )^{1/3}.
\end{eqnarray}
From these expressions we see that for $\Lambda > 0$ there is
cosmological horizon.
In  FIG.~(\ref{fig:flat02}) we show different situations for
negative $\Lambda < 0$. Firstly, the solid line represent an
extreme black hole (where we have used the values $M=7.268$,
$\Lambda=-2$ and $q=2$). Secondly, the dashed line represents a
naked singularity (with values $M=1$, $\Lambda=-2$ and $q=2$.
Finally, the dotted line correspond to a black hole with two
horizons (here, $M=25.15$, $\Lambda=-2$ and $q=2$). When $q=0$ we
get $r_{-}=0$ and $r_{+}=\sqrt{-M/\Lambda}$, which are the
horizons of the  anti de Sitter 2+1 metric.

An interesting case results  when $\Lambda = 0$. Here, we have a
charged  solution with a cosmological horizon at
\begin{eqnarray}
r_{_{H}}= \frac{4 q^2}{3 M}.
\end{eqnarray}
Notice that the gravitational field is asymptotically flat in this
case. As far as we know, there are no other solutions with
cosmological horizon which exhibits the same property.

In order to study the thermodynamics of our solutions, we evaluate
the temperature of the black hole, which is given in terms of its
surface gravity~\cite{Brown2,Visser}
\begin{eqnarray}
\label{T}
k_{_{B}} T_{_{H}} = \frac{\hbar}{2 \pi} \, k.
\end{eqnarray}
For a circularly symmetric metric  the surface gravity can be
computed via \cite{Hawking,Romans}
\begin{eqnarray}
\label{k} k=\frac{1}{2}\left | \lim_{r \rightarrow r_{_{+}}}
\frac{\partial_{r} g_{tt}}{\sqrt{- g_{tt} g_{rr}}}  \right |,
\end{eqnarray}
where $r_{_{+}}$ is the event horizon. From our solution and
$\Lambda = - 1/l^2 < 0$ we get that
\begin{eqnarray}
\label{temp} k_{_{B}} T  = \frac{\hbar}{4 \pi} \hspace{5cm}
\nonumber
\\ \times \left | - \sqrt{\frac{16 M}{3 l^2}} \cos \, \left (
\frac{1}{3} \, \arccos \left [ \frac{2 q^2}{\sqrt{ \frac{M^3
l^2}{3}}} \right ] + \frac{2 \pi}{3}\right ) \right . \nonumber \\
\left . - \frac{q^2}{M l^2} \cos^{-2} \, \left ( \frac{1}{3} \,
\arccos \left [ \frac{2 q^2}{\sqrt{\frac{M^3 l^2}{3}}} \right ]  +
\frac{2 \pi}{3}\right ) \right | \,\,\,\,\,\,\,\,\,.
\end{eqnarray}
It is easy to check that when $q=0$, $T$  reduces to the static
BTZ temperature~\cite{BaTeZa}. In the extreme case~(\ref{rextr}),
the temperature vanishes.

If $\Lambda=0$ the temperature at the cosmological horizon becomes
$T= \frac{3 \hbar}{16 \pi} \frac{M^2}{q^2}$.

The entropy of the black hole can be trivially obtained by using
$S= 4 \pi r_{_{+}}$. Other thermodynamic quantities such as heat
capacity and chemical potential can be computed as
well~\cite{Brown2}. We do not discuss them here.

Our solution is singular only at $r=0$. Effectively, the first two
invariant curvature scalars are
\begin{eqnarray}
R= 6 \Lambda
\end{eqnarray}
\begin{eqnarray}
R_{ab}R^{ab}=12 \Lambda^2 + \frac{8 q^4}{3 r^6}.
\end{eqnarray}
As we mentioned above, there is a true singularity  at the origin
of the coordinate system.  Notice that these invariants are
nonsingular at the horizons.

In this paper we have investigated nonlinear electrodynamics
restricted to traceless energy momentum in 2+1 dimensions. The
obtained solutions for a circularly symmetric metric describe
black holes with a Coulomb-like field, which are asymptotically
anti-de Sitter spacetimes. It is worthwhile to point out that the
derived charged black holes possess finite mass contrary to the
charged BTZ solutions, which because of the presence of the
logarithmic term in the metric yields a divergent quasilocal mass.
The solution with $\Lambda = 0$ is remarkable since it has a
cosmological horizon and the spacetime is asymptotically flat.

M.C. was supported by FONDECYT (Chile) under Grant \# 1990601 and
Direcci\'{o}n de Promoci\'{o}n y Desarrollo de la Universidad del
B\'{\i}o-B\'{\i}o, N.C  by USACH-DICYT (Universidad de Santiago de
Chile) under Grant \# 0497-31CM, S.d.C. was supported by FONDECYT
(Chile) under Grant \# 1000305 and by UCV-DGI Grant \# 123.744/00
and A.G.was supported by CONACYT (M\'exico)by grant 32138E.

\end{document}